\documentclass[letterpaper]{article}
\usepackage{aaai}
\usepackage{times}
\usepackage{helvet}
\usepackage{courier}
\begin{document}

\title{Studying Politically Vulnerable Communities Online: \\ Ethical Dilemmas, Questions, and Solutions}
\author{Robert Gorwa \quad 
Philip N. Howard \\ Oxford Internet Institute, University of Oxford}
\maketitle
\begin{abstract}

This short article introduces the concept of political vulnerability for social media researchers. How are traditional notions of harm challenged by research subjects in politically vulnerable communities? Through a selection of case studies, we explore some of the trade-offs, challenges, and questions raised by research that seeks be  robust and transparent while also preserving anonymity and privacy, especially in high--stakes, politically fraught contexts.

\end{abstract}

\section{1. Introduction}
\noindent As studying social media has become a major cross-disciplinary endeavor in the past decade, scholars have grappled with the numerous  ethical and methodological questions that Internet researchers inevitably face on a daily basis. In the significant body of work about online research ethics, which touches on everything from anonymization to the researcher-subject relationship, relatively little has been written about notions of vulnerability. Who are vulnerable subjects in online research? Interestingly, Internet researchers seem to generally conceptualize vulnerability in the biomedical research tradition --- focusing on children, for instance, or other populations unable to fully provide informed consent --- as opposed to other forms of vulnerability, especially that which we term \textit{political vulnerability}. 

In this short paper, we discuss the methodological and ethical quandaries that can emerge when studying politically vulnerable communities, such as such as political dissidents, bloggers in authoritarian countries, refugees, and others. Our central question is as follows: to what extent are current social media research guidelines acceptable for studying these types of communities, or should researchers go beyond current ethics standards to treat them with a unique set of principles? 

The first section of the paper will engage with the literature on Internet ethics in order to (a) outline the central concepts for ethical online research and (b) see how scholars of Internet research ethics conceptualize the notion of the ``vulnerable subject" or ``vulnerable community." The second section will aim to contextualize this literature with a number of short case studies (relevant journal articles and reports), to demonstrate whether or not the best practices and ethics guidelines discussed in the first section are generally applied by researchers. At the ICWSM workshop, we hope to build on these two sections by discussing alternative approaches for minimizing harm in online research that deals with politically vulnerable communities. A preliminary discussion is presented in the final section. 

\section{2. Vulnerability and Vulnerable Communities}
Internet researchers generally accept that certain populations are often more vulnerable in all types of research, whether offline or online, and that some sort of duty to protect those subjects exists, depending on the circumstances (Markham and Buchanan, 2012). While the definition of vulnerability presented by scholars of Internet ethics is often broad, they tend to define ``vulnerable persons" in the biomedical research tradition as ``young people, the elderly, and people with mental health issues" (Markham and Buchanan, 2012: 29). Buchanan (2011: 85) provides an example of this reasoning when she offers an example of vulnerable populations in Internet research that includes pregnant women and prisoners. These are persons vulnerable in the traditional sense of human-subject research: people who may not be able to provide properly informed consent, or could be harmed by direct experimentation. However, many  communities exist online that are vulnerable in a different sense, wherein their information must remain private because their identification could be problematic in some way. For example, Bruckman (2002) classifies LGBTQ+ individuals and those with serious diseases as vulnerable populations that could experience significant harm if ``outed" and de-anonymised. Her work presents a broader notion of vulnerability, where the issue is not with just with subjects being unable to provide truly informed consent, but also more broadly with the sensitive nature of the data being collected by the researcher (and therefore, with the practices that the researcher takes to protect their subjects). 

We would like to introduce a notion of political vulnerability, where research subjects are vulnerable for reasons that go beyond them simply not being able to provide properly informed consent, and extend specifically into harms that could arise due to their political and social context. For example, in many studies, especially those conducted on the ``authoritarian Internet,''
data collected by the researcher could prove politically damning and harmful to the research subjects. Interestingly, direct mentions of political vulnerability seem to be entirely absent from the Internet ethics literature, and one must look towards the conventional ethics literature to find allusions to this concept (Birman, 2006) --- a potentially concerning gap given the amount of research currently being done on politically vulnerable communities online.

However, although the current notions of vulnerability commonly discussed in the Internet ethics community can seem a bit narrow, the general principles espoused by the Internet ethics community are far broader. The first guiding principle established by the Association of Internet Researchers in 2012 actually concerns vulnerability, stating that ``The greater the vulnerability of the community / author / participant, the greater the obligation of the researcher to protect the community / author / participant." This flexible principle would allow for the researcher to protect politically vulnerable communities as long as they properly identify the community as such. Approaches which emphasise context are crucial, as several factors should be taken into account when dealing with politically vulnerable communities, such as political dissidents, that may not be as important in other scenarios. Hongladarom and Ess (2007) have proposed a set of ``good samaritan ethics" which could provide a good starting point for those studying politically vulnerable subjects. They advocate a set of best practices that hold for both qualitative and quantitative work, and include: removing all names and pseudonyms from the published work, only accessing publicly available content, not sneaking into gated online forums or communities, identifying oneself as a researcher rather than masquerading as a user, and not linking to direct sources in the published work to help maintain anonymity. Leaving aside the point that even without links it may takes a quick copy-and-paste online search to find quotations, provided that they have been reproduced in their language of origin, this set of principles raises several methodological questions which apply to most approaches that deal with vulnerable subjects in online research. 

The central issue is one of credibility and trustworthiness. As Bruckman (2002) has noted, the more a researcher protects sources and anonymizes, the more that they may be seen as reducing the accuracy and replicability of their study. This may be more of an issue for quantitative, as opposed to qualitative scholars (whose work is generally not designed to be reproducible), but a overall assumption seems to be that there can be a trade off between the robustness of one’s research and the amount of protection granted to one’s subjects. How best to ensure that one’s findings are reliable, while one’s sources remain safely protected? Eynon, Fry, and Schroeder (2008: 35) have argued that a key question relates to ``the extent to which Internet researchers should be concerned with the collection and use of potentially harmful data --- given that [they] cannot anticipate all the ways in which it might be reused and by whom." If, in the case of a politically vulnerable community, the Internet researcher is very concerned about the protection of data, and yet still wants to maintain methodological rigour, how should they best proceed? In the following section, we shall present some brief case studies to evaluate how scholars deal with this problem.

\section{3. Case Studies}
Curious to understand whether or not the guidelines and ethical frameworks proposed by Internet ethics scholars are broadly employed in actual research, we collected a convenience sample of 15 articles on the topic of online political debate in authoritarian countries. Using the Scopus database aggregator to search for articles with the boolean search function ``political debate AND online AND authoritarian," we selected a diverse set of articles published between 2000-2015, which employ a variety of methods --- social media scraping, surveys, interviews, ethnography, participant observation, discourse analysis, and more --- and deal with five countries: China, Iran, Burma, Cuba, and Uzbekistan. Some articles are specifically about blogging, while others are predominantly concerned with forums, micro-blogs, and other forms of social media. Given the length constraints of this paper, we will present what we believe to be the most interesting findings, rather than discuss each article individually in depth.

In their article on ``News Blogging in Cross-Cultural Contexts," Katz and Lai (2009) provide an interesting method for protecting source material. They seek to understand the motives of bloggers in South-East Asian Countries, especially in countries such as Burma and Thailand where speech is restricted and bloggers can be jailed or physically punished for their online activities. Instead of simply quoting online content, which, as noted earlier, can easily be de-anonymized through a cursory Google search, they reach out to interesting bloggers and interview them via e-mail. The authors can then publish anonymous, original quotations which provide interesting insights into blogging/online dissent as a practice, without providing searchable, potentially incriminating material. The authors do not discuss their email procedure in great detail, but this approach seems to be a good option with the caveat that the researchers (and the bloggers) must maintain an adequate level of operational security, and ensure that they use encrypted email (PGP) and secure communications channels. 

This research design stands in contrast to the technique employed by Yang (2003), who is interested in debates around democratization on Chinese social media. Yang performs a discourse analysis of various Chinese forums and bulletin boards, and reproduces many posts from these forums in order to illustrate his main arguments. He provides usernames and the post dates for each quote, in order to ``give electronic voice to those posting in the forum" (Yang, 2003: 416). This technique could be perceived as problematic from an ethical standpoint, especially since most of the quotes reproduced are of users critiquing the government and proclaiming the importance of free speech above all else. It is interesting to note his rationale --- he not only attaches these identifiers for citation purposes (in what he surely considers to be a transparent and robust practice), but also to credit the authors and give them a ``voice." While Yang notes that government censors are very active on these sites (and, ostensibly, posts which are deemed unsuitable by the government will simply be removed), he does not mention the potential implications of capturing posts before they are censored, or that publishing them and linking them back to a user could provide them with more voice than they may have wished. He also does not mention whether or not these forums are fully private or not, and whether or not a site registration was required to gain access this material. If so, social media research guidelines generally suggest that some sort of consent and moderator permission should be required (Markham and Buchanan, 2012). Shen (2009) employs similar approach. He provides exact (translated) quotes taken from forums and bulletin boards, and then quotes the reference numbers and usernames for each post — again, a potentially problematic practice.

In an influential article, King, Pan, and Roberts (2013) perform a large scale quantitative analysis of Chinese censorship online. Their highly publicized finding was that the Chinese government does not necessarily censor commentary critical of the government, as commonly believed, but rather predominantly censors posts that advocate for some sort of collective action. They anonymize all of the social media content used to corroborate their argument; however, when presenting quotes, they choose to provide the original Chinese posts alongside an English translation. While this approach allows greater transparency  as far as findings go (given the amount of flexibility that is inherent in the act of translation, authors could unconsciously or consciously provide subtle nudges to the translations in order to better support their argument), it also significantly increases the chances that the author of the quote can be determined via a Chinese  search engine. This could be problematic, especially since many of the posts they choose to publish are quite inflammatory and highly critical of the government, and because they know that several of these posts were eventually censored. They even go as far as to comment that these posts constitute ``detailed information that the Chinese government does not want anyone to see and goes to great lengths to prevent anyone from accessing" (King, Pan, and Roberts, 2013: 328), and yet they do not discuss whether reproducing this information could perhaps endanger the (possibly unwitting) participants of this study. 

Kelly and Etling's study of Iranian blogging (2008), published by Harvard's Berkman Center, provides a potential response to the transparency-vulnerability trade-off described above. They conduct a network analysis of more than 60 000 blogs in order to map key political issues in the Iranian blogosphere, and employ a team of Iranian language coders, who assess important blogs individually and provide short summaries of their content in English. Rather than provide quotes directly from the blogs, or linking to the blogs themselves, the authors choose to publish these summaries. For example, one reads ``Blogger believes that Iran lacks basic freedoms and democracy and posts articles, poems, and pictures to reflect his beliefs." (Kelly and Etling, 2008: 14). These brief quotes are used to drive the analysis, and provide an admirable level of anonymity and security to the potentially vulnerable bloggers who are being studied. However, the authors do not employ any tests for intercoder reliability (where multiple coders code the same randomly sampled content, so that the authors can ensure that they are all similar), potentially raising questions about the robustness of the study. Because it appears that the authors themselves do not speak Farsi, it is possible that coders could affect the study by coding erroneously (either accidentally, or on purpose) without the authors realizing it. 

Two studies of political dissidents and bloggers online, by Venegas (2010) and Kedzior (2011) utilize a similar technique in order to maintain reliability and address the vulnerability of their sources. In Venegas' study of Cuban dissidents, and Kedzior's study of Uzbek exiles, both scholars only cite material that is produced by highly visible political dissidents/bloggers/intellectuals. Kedzior focuses on the online forums and communities through which Uzbek diaspora communities engage politically with their counterparts still in Uzbekistan, and punctuates her account with the online writings of Xoldor Vulqon, a high-profile poet and political commentator. Similarly, Venegas’ critical analysis of the ``biopolitics of Cuban blogging" only quotes material written by well known Cuban bloggers/intellectuals, such as Yoani Sanchez (an award winning blogger and writer). While neither author explicitly acknowledges that this technique is intended to protect the identities of other, less public bloggers, it may provide an effective solution to the vulnerability problem. The authors do not have to de-anonymize their sources, as their subjects are already very public and well known, and therefore, have in effect personally absorbed the risks associated with their dissent. It is not to say that these risks do not exist --- Vanegas provides several examples of bloggers describing, in excruciating detail, the many threats levied against them by the state  --- but the calculus of ethical vulnerability is slightly different, as the threshold of harm that could possibly face  the research subject is unlikely to be raised by a published piece of research, given their high public profile and history of public dissent. 

Taken together, these brief examples provide some insight into the ways that researchers deal with the political vulnerability of their data and their subjects. In the next section, we will draw upon these examples to inform a discussion of best practices for future research in this area.

\section{4. Discussion}
What is the best way to maintain methodological rigor without sacrificing the safety of potentially politically vulnerable subjects?  The examples provided above implicitly hint at a trade-off between robustness and the level of protection granted to sources. After all, in a perfect world, scholars would be able to reproduce all of the content that they are using to corroborate their argument, and would be able to release their data publicity to maximize transparency (and the validity of their study). However, in the case of politically vulnerable communities or politically sensitive material, this is clearly not possible, and many of the countries discussed in the previous section --- China, Burma, Iran --- have a history of imprisoning or physically punishing those who cross the line with their speech online. As such, the standard of harm for the research subjects is often much higher than in North America or Europe, and scholars performing online research in these areas should be cognizant about the potential vulnerability of their subjects.  

Interestingly, although none of the authors mentioned above directly discuss the potential ethical issues posed by their study, they seem to have an implicit understanding that their subjects may be vulnerable, and they operationalize these assumptions at least to a certain extent. For example, King, Pan, and Roberts assert the need to maintain anonymity for their data, and state that releasing their full dataset would constitute an ethical breach. However, as mentioned in the previous section, they still reproduce full quotations in a way which could be problematic. Since we can assume the good intentions of most researchers, and also assume that every scholar wishes for their work to be as robust as possible, what are the best ways to move beyond this trade-off? How can we ensure that research is robust, and yet still adequately protects its potentially politically vulnerable subjects?

When discussing best practices for this sort of research in the future, perhaps the first step would be to propose that scholars make an effort to be more open about the potential political vulnerability of their subjects, and that they discuss these issues in the methodology sections of their articles. A second step would be to push for creative solutions to the searchability problem --- the issue that direct quotations can possibly be traced back to the author even if they are anonymized. Katz and Lai's interview method provides one such solution: they conduct online interviews with their subjects, instead of quoting online content. As long as they maintain sufficient operational security, and take care to use encrypted email and secure their communications well, it will be far more difficult to de-anonymize those quoted. This method is also fairly robust as it allows them to provide direct quotations, although the sources must, of course, remain anonymous. The technique employed by Vanegas and Kendzior is also good, where they only quote high-profile bloggers and public intellectuals. These individuals, while not necessarily less vulnerable, are less likely to face harm as a direct result of the research, given their existing levels of high-profile public dissent.  

Kelly and Etling's method of summarizing the content, rather than quoting it explicitly, could provide another acceptable solution, as long as intercoder reliability is maintained. As well, some sort of third-party review could be introduced, where other trusted academics or researchers could assess qualitative data and vouch for its validity (to thwart, for example, the potential scenario that journal editors worry about, involving the fabrication of content by unscrupulous researchers), thereby increasing robustness while maintaining high levels of anonymity. Their technique is taken a step further by Markham (2012: 334), who argues that the only way to truly ensure the privacy of the source material is through ``creative, bricolage-style transfiguration of original data into composite accounts or representational interactions." While this method is very well suited for studying politically vulnerable communities, according to Markham, several papers written by herself and her colleagues were rejected on the grounds that they had ``fabricated" data, while they were really trying to protect their sensitive source material. This example illustrates that both qualitative and quantitative scholars may face institutional pressures to make at least certain aspects of their data public or ``transparent," at a potential cost to the safety of their research subjects. 

In sum, these are only some possible solutions to the robustness-anonymization trade-off. Social media  researchers must emphasize that context is key, and that the best solutions will have to be determined on a case-by-case basis. It should be noted that not all these best practices will be acceptable for those with a more positivist bent, while post-positivists are more likely to employ more involved techniques such as the one proposed by Markham. Also, both qualitative and quantitative scholars should find ways to account for political vulnerability, although the way this is accomplished may vary. The notion of what constitutes robust research can vary between qualitative and qualitative approaches, with quantitative scholars in certain disciplines (such as Psychology or Political Science) more likely to emphasize the importance of reproducibility. However, we would suggest that these short case studies illustrate that it is completely possible to produce robust work which also accounts for political vulnerability: in the small number of case studies provided here, we have seen several creative solutions. Further discussion and experimentation in this are would surely demonstrate that many others are possible as well, depending on the context and subject population in question.

\section{5. Conclusion}
The Internet ethics community has spent almost two decades discussing critical ethical questions associated with Internet research. In this paper, we have suggested that the notion of \textit{political vulnerability} has been underrepresented in these debates, and conclude that it would be worthwhile for researchers to (a) take extra care when working with politically vulnerable communities, and (b) be cognizant that their work could have direct ramifications on the politically sensitive persons who (often unwittingly) provide data. Given that Internet researchers emphasize the importance of making informed methodological decisions based on context, it would be valuable for them to ensure that they consider \textit{political context} as well. A credible research method should not only be methodologically rigorous and well suited to the task at hand, but also be ethically credible and robust. It should demonstrate that the researcher has treated their data (and the populations that could be affected by said data) with the appropriate measure of care, and that they have thought through some of the issues associated with anonymity presented above. After all, no researcher wants to be informed that someone has been physically harmed as a result of their study. In order to avoid this possibility, responsible researchers would do well to strive towards best practices for research that deals with politically vulnerable populations.

\section{6. References}

\smallskip \noindent Bassett, E, and O'Riordan, K. 2002. Ethics of Internet Research: Contesting the Human Subjects Research Model. \textit{Ethics and Information Technology} 4 (3): 233--247.

\smallskip \noindent Birman, D. 2006. Ethical Issues in Research With Immigrants and Refugees. In The \textit{Handbook of Ethical Research with Ethnocultural Populations and Communities}, Trimble and  Fisher, eds. Thousand Oaks, California: SAGE.

\smallskip \noindent Bruckman, A. 2002. Studying the Amateur Artist: A Perspective on Disguising Data Collected in Human Subjects Research on the Internet. \textit{Ethics and Information Technology} 4 (3): 217--231.

\smallskip \noindent Buchanan, E. 2011. Internet Research Ethics: Past, Present, and Future. In \textit{The Handbook of Internet Studies}, Consalvo and Ess, eds. 83--108. Oxford, UK: Wiley-Blackwell.

\smallskip \noindent Elgesem, D. 2002. What Is Special about the Ethical Issues in Online Research? \textit{Ethics and Information Technology} 4 (3): 195.

\smallskip \noindent Ess, C., and the Association of Internet Researchers Working Group. 2002. Ethical Decision-Making and Internet Research: Recommendations from the AoIR Ethics Working Committee. 

\smallskip \noindent Eynon, R., Fry, J., and Schroeder, R. 2008. The Ethics of Internet Research. \textit{In The Handbook of Online Research Methods}, edited by Blank, Lee, and  Fielding, 23--41. London, UK: SAGE.

\smallskip \noindent Eynon, R., Schroeder, R., and Fry, J. 2009. New Techniques in Online Research: Challenges for Research Ethics. \textit{Twenty-First Century Society} 4 (2): 187--199.

\smallskip \noindent Hongladarom, S., and Ess, C., eds. 2007. \textit{Information Technology Ethics: Cultural Perspectives}. Hershey, PA: Idea Group.

\smallskip \noindent Katz, J., and Lai, C.H. 2009. News Blogging in Cross-Cultural Contexts: A Report on the Struggle for Voice. \textit{Knowledge, Technology and Policy} 22 (2): 95--107.

\smallskip \noindent Kelly, J., and Etling, B. 2008. Mapping Iran’s Online Public: Politics and Culture in the 	Persian Blogosphere. Berkman Center Research Report.

\smallskip \noindent King, G., Pan, J., and Roberts, M.E. 2013. How Censorship in China Allows Government Criticism but Silences Collective Expression. \textit{American Political Science Review} 107 (2): 326--343.

\smallskip \noindent Kendzior, S. 2011. Digital Distrust: Uzbek Cynicism and Solidarity in the Internet Age. \textit{American Ethnologist} 38 (3): 559--575.

\smallskip \noindent Markham, A., and Buchanan, E. 2012. Ethical Decision-Making and Internet Research: Recommendations from the AoIR Ethics Working Committee (Version 2.0).

\smallskip \noindent Markham, A. 2012. Fabrication as Ethical Practice: Qualitative Inquiry in Ambiguous Internet Contexts. \textit{Information, Communication and Society} 15 (3): 334--353.

\smallskip \noindent Schroeder, R. 2007. An Overview of Ethical and Social Issues in Shared Virtual Environments. \textit{Futures} 39 (6): 704–-717.

\smallskip \noindent Shen, S. 2009. A Constructed (un)reality on China's Re-Entry into Africa: The Chinese Online Community Perception of Africa. \textit{Journal of Modern Africa Studies} 47 (3): 425--448.

\smallskip \noindent Venegas, C. 2010. `Liberating' the Self. \textit{The Journal of International Communication} 16 (2): 43--54

\smallskip \noindent Walther, J.B. 2002. Research Ethics in Internet-Enabled Research: Human Subjects Issues 	and Methodological Myopia. \textit{Ethics and Information Technology} 4 (3): 205.

\smallskip \noindent White, M. 2002. Representations or People? \textit{Ethics and Information Technology} 4 (3): 249--266.

\smallskip \noindent Yang, G. 2003. The Co--Evolution of the Internet and Civil Society in China. \textit{Asian Survey} 43 (3): 405--422.

\end{document}